\documentclass[final,copyright,creativecommons]{eptcs} %

\usepackage{csquotes}
\usepackage[british]{babel} %
\usepackage{graphicx}
\usepackage{wrapfig}
\usepackage{tabularx}
\usepackage{booktabs}
\usepackage[normalem]{ulem}
\usepackage{amsmath}
\usepackage{amssymb}
\usepackage{subcaption}
\usepackage{multicol}

\usepackage[T1]{fontenc}

\usepackage{amsthm}
\theoremstyle{definition}
\newtheorem{definition}{Definition}[section]

\usepackage[dvipsnames]{xcolor} 
\definecolor{codegreen}{rgb}{0,0.6,0}
\definecolor{codegray}{rgb}{0.5,0.5,0.5}
\definecolor{codego}{rgb}{0.9,0.9,0.85}
\definecolor{codesql}{rgb}{0.9,0.85,0.9}
\definecolor{codesparql}{rgb}{0.9,0.85,0.85}
\definecolor{codepurple}{rgb}{0.58,0,0.82}
\definecolor{backcolour}{rgb}{0.95,0.95,0.92}

\usepackage{listings} 
\lstdefinelanguage{prism}{
  morekeywords={
    A, bool, clock, const,
    ctmc, C, double, dtmc, E, endinit, endinvariant, endmodule, endrewards, endsystem,
    false, formula, filter, func, F, global, G, init, invariant, I, int, label, max, mdp,
    min, module, X, nondeterministic, Pmax, Pmin, P, probabilistic, prob, pta, rate,
    rewards, Rmax, Rmin, R, round, S, stochastic, system, true, U, W,
    observables, endobservables, observable
  },
  morecomment=[l]{//}}
\lstdefinestyle{prism}{
    mathescape=true,
    backgroundcolor=\color{backcolour},
    basicstyle=\ttfamily\footnotesize\color{red},
    commentstyle=\color{codegreen},
    keywordstyle=\color{black}\bfseries,
    identifierstyle=\color{red}\slshape,
    numberstyle=\tiny\color{codegray},
    stringstyle=\color{codepurple},
    emphstyle=\color{blue},
    morestring=[b]",
    breakatwhitespace=false,
    breaklines=true,
    columns=flexible,
    captionpos=b,
    keepspaces=true,
    numbers=left,
    numbersep=2pt,
    xleftmargin=8pt,
    showspaces=false,
    showstringspaces=false,
    showtabs=false,
    tabsize=4
}
\usepackage{cleveref}
\Crefname{lstlisting}{Listing}{Listings}
\crefname{sublstlisting}{listing}{listings}
\Crefname{sublstlisting}{Listing}{Listings}
\AtBeginDocument{\DeclareCaptionSubType{lstlisting}}
\usepackage{soul} 
\usepackage{xspace,extdash,underscore} %

\newcommand\mytexttt[1]{\begingroup\sethlcolor{backcolour}\hl{~\texttt{#1}~}\endgroup} 
  
\newcommand\prismid[1]{\textcolor{red}{\texttt{#1}}}
\newcommand\tlop[1]{\mathbf{#1}}

\providecommand{\event}{FMAS 2024}
\author{Till Schnittka and Mario Gleirscher
  \institute{University of Bremen, Germany}
  \email{\{schnitti,gleirsch\}@uni-bremen.de}}
\date{\today}
\title{Synthesising Robust Controllers for Robot Collectives\\
  with Recurrent Tasks: A Case Study}
\def\titlerunning{Robustly Recurrent Strategies for Robot Collectives}
\def\authorrunning{Till Schnittka and Mario Gleirscher}

\makeatletter
\begin{document}
\maketitle

\begin{abstract}
  When designing correct-by{\Hyphdash}construction controllers for
  autonomous collectives, three key challenges are the task
  specification, the modelling, and its use at practical scale.  In
  this paper, we focus on a simple yet useful abstraction for
  high-level controller synthesis for robot collectives with
  optimisation goals~(e.g.,\xspace maximum cleanliness, minimum energy
  consumption) and recurrence~(e.g.,\xspace re-establish contamination
  and charge thresholds) and safety~(e.g.,\xspace avoid full discharge, mutually
  exclusive room occupation) constraints.
  Due to technical limitations~(related to scalability and using
  constraints in the synthesis), we simplify our graph-based setting
  from a stochastic two-player game into a single-player game on a
  partially observable Markov decision process~(POMDP).  Robustness against environmental uncertainty is
  encoded via partial observability.  Linear-time correctness
  properties are verified separately after synthesising the POMDP
  strategy.
  We contribute at-scale guidance on POMDP modelling and
  controller synthesis for tasked robot collectives exemplified by the
  scenario of battery-driven robots responsible for cleaning public
  buildings with utilisation constraints.
\end{abstract}

\section{Introduction}
\label{sec:orgb3d576a}

Hygiene in public buildings has been hotly debated ever since the
increased safety requirements during the coronavirus pandemic.
Autonomous robot collectives can help to relieve cleaning staff and
keep highly frequented buildings (e.g.,\xspace hospitals, schools) clean.
However, commercially available solutions have their limitations, for
example, a need for manual task programming or a lack of online
adaptability.  In contrast to domestic homes, there are strict
regulations (e.g.,\xspace \cite{uba} for schools) that stipulate which areas
must be cleaned and at what intervals.
Changing operational conditions (e.g.,\xspace room occupation, equipment
reconfiguration, cleaning profile, regulations) create the need for an
automatic generation of cleaning schedules for robot collectives and
for proving their compliance with hygiene requirements.
This scenario is an instance of a multi-faceted \emph{recurrent
  scheduling} problem discussed below.

\paragraph{Cleaning Buildings as a Running Example.}
\label{sec:org5c08ef7}

\begin{figure}
  \centering
  \subcaptionbox{Room plan\label{room-layout}}[.45\linewidth]%
  {\includegraphics[height=3cm,trim=10 10 5 5,clip]{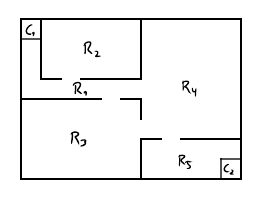}}
  \subcaptionbox{Room utilisation plan\label{room-times}}[.45\linewidth]{%
    \small
    \begin{tabular}{ll}
      \toprule
      \textbf{Room} & \textbf{Time of use}\\
      \midrule
      \(\mathcal{R}_1\) & 8:00 -- 10:00\\
      \(\mathcal{R}_2\) & 12:00 -- 14:00\\
      \(\mathcal{R}_3\) & 10:00 -- 12:00\\
      \(\mathcal{R}_4\) & 15:00 -- 16:00\\
      \bottomrule
    \end{tabular}}
  \caption{Examples of a room plan and the per-room utilisation}
\end{figure}

When planning the cleaning of a building~(e.g.,\xspace a school), we can use
its layout in the form of a room plan (\Cref{room-layout}).  Apart
from a set of $m$ rooms $\mathcal{R}=\{\mathcal{R}_1, \ldots, \mathcal{R}_m\}$ with an assigned
area, such plans include the connections between rooms and the
locations of $n$ charging stations $C=\{C_1, \ldots, C_n\}$.  A
collective of $k\leq n$ robots $B=\{B_1,\ldots,B_k\}$ is responsible
for cleaning~$\mathcal{R}$.  $B$ is tasked to keep $\mathcal{R}$ clean while charging
its batteries using~$C$.  Each robot has a limited battery size and a
charging point in $C$ as an assigned resting position.
Additionally, there is a room utilisation plan (\Cref{room-times})
containing the times when the rooms are in use, which can change on a
daily basis.  While a room is in use, no cleaning robot may be inside
and, thus, cannot clean it.

The task is to create a cleaning \emph{strategy} (or schedule) for
$B$, constrained by the utilisation plan and charging needs.
Moreover, this strategy should keep the rooms clean enough, while
minimising battery consumption.
To specify this task adequately, we define cleanliness in terms of
\emph{contamination}.  Since we are dealing with floor cleaning
robots, we limit our definition to floor surfaces.  For example,
hygiene guidelines for schools~\cite{uba} and the associated standard
DIN\,77400~\cite{DIN77400}
recommend cleaning intervals for floor surfaces.  We reflect this
recommendation in our definition by \emph{contamination rates and
  thresholds}.  Each room has a certain contamination rate.  Over time,
total contamination accumulates in a room and can be reset by cleaning.
Clearly, the total contamination of a room should not exceed a certain
threshold.

\paragraph{Approach.}

We propose a quantitative stochastic approach to synthesise strategies
for robot collectives with recurrent tasks, such that the strategies
are robustly (i.e.,\xspace under uncertainty) compliant to recurrence and
safety constraints and optimisation goals.
We consider (i) weighted stochastic models and strategy synthesis for
(ii) optimally coordinating robot collectives while (iii) providing
guarantees (e.g.,\xspace recurrence, safety) on the resulting strategies (iv)
under uncertainty (e.g.,\xspace partial observability).  We select
POMDPs, a generalisation of Markov decision processes~(MDPs), for our problem.

\paragraph{Related Work.}
\label{sec:orgf01d088}

Among the works employing POMDPs for optimal planning,
Macindoe et al.~\cite{Macindoe2012-POMCoPBeliefSpace} show strategy synthesis for
human-robot cooperative pursuit games with robots and humans acting in
turns.  Moreover, Thomas et al.~\cite{Thomas2021-MPTPMotionplanning} combine
PDDL-based high-level task scheduling and POMDP-based low-level
navigation.  They use a Kalman filter to predict the distribution of
the POMDP's belief state based on discretised robot dynamics.

Using the \textsc{Prism}\xspace model checker, Giaquinta et al.~\cite{auto} show the synthesis of
minimal-energy strategies for robots finding fixed objects.  Object
finding as an instance of navigation can be solved with memory-less
strategies, that is, functions of the current state (here, positions
of robot and object).
Lacerda et al.~\cite{Lacerda2017-MultiObjectivePolicy} propose
multi{\Hyphdash}objective synthesis of MDP policies satisfying
\emph{bounded co-safe LTL} properties using \textsc{Prism}\xspace.  Illustrated by a
care robot, they employ a timed
MDP %
and filter irrelevant transitions and states, resulting in a
reduced %
MDP where time as a state variable preserves bounded properties.
Basile et al.~\cite{Basile2020-StrategySynthesisAutonomous} use stochastic
priced timed games and reinforcement learning~(RL)-enhanced statistical model checking
(with \textsc{Uppaal} Stratego) to synthesise \emph{safe,
  goal-reaching, and minimal-arrival-time} strategies for a
\emph{single} autonomous train operated under moving-block signalling.

El Mqirmi et al.~\cite{Mqirmi2021-AbstractionbasedMethod} combine multi-agent
RL and verification to coordinate robot collectives.  An abstract
MDP is generated from expert knowledge for optimal synthesis of a
joint abstract strategy (using \textsc{Prism}\xspace, \textsc{Storm}) under PCTL
(safety) constraints.  RL identifies a concrete strategy within
these constraints by using shielding (i.e.,\xspace only choosing actions
compliant with the abstract strategy).
Gu et al.~\cite{Gu2022-Correctnessguaranteedstrategy} tackle state space
reduction via RL to synthesise optimal %
navigation and task schedules for collectives (e.g.,\xspace a quarry
with autonomous vehicles) and timed games (in a \textsc{Uppaal}
Stratego extension) to check timed CTL properties (e.g.,\xspace
\emph{liveness, safety, reachability}) of the synthesised strategies.

V{\'a}zquez et al.~\cite{Vazquez2022-SchedulingMissionsConstrained} developed a
domain-specific language~(DSL) for specifying tasks for collectives.  Task allocation
constraints are solved by \textsc{Alloy} and plain MDPs are
employed (via \textsc{Prism}\xspace, \textsc{EvoChecker}) to synthesise
\emph{goal-reaching, minimum-travel-time} schedules.

\paragraph{Contributions.}

Our approach enhances works in optimal
planning~\cite{Macindoe2012-POMCoPBeliefSpace,
  Thomas2021-MPTPMotionplanning} by a step of strategy verification
against stochastic temporal properties.
Object finding~\cite{auto} corresponds to a \emph{reachability}
property, whereas continuous contamination and its removal to a
\emph{response} property.  Moreover, object finding differs from
\emph{recurrent scheduling} in that it is static and can be realised
with a simpler reward function and a smaller state space.  Our problem
involves a dynamic goal %
with robots operating 24~hours a day, having to coordinate their work
continuously.
Beyond bounded co-safe LTL~\cite{Lacerda2017-MultiObjectivePolicy},
our approach supports \emph{bounded response} properties
$\tlop{G}(\phi\to\tlop{F}^{\leq T}\psi)$.
The above works~\cite{Lacerda2017-MultiObjectivePolicy,
  Basile2020-StrategySynthesisAutonomous,
  Mqirmi2021-AbstractionbasedMethod,
  Gu2022-Correctnessguaranteedstrategy} underpin the usefulness of
stochastic abstractions for synthesis in various domains.  Our use of
POMDPs to hide parts of the stochastic process (e.g.,\xspace
contamination) and explicit concurrency (e.g.,\xspace for many simultaneous
robot movements) offers an alternative to obtaining small models for
multi-agent synthesis under limited resources (e.g.,\xspace time constraints
to clean rooms).
Additionally, we argue how the model of our case study---a collection
of cleaning robots subjected to hygiene requirements---, while kept
simple for illustrative purposes, scales and generalises to a range of
similar scenarios in other application domains.
Apart from our focus on recurrence and model reduction, a
DSL~\cite{Vazquez2022-SchedulingMissionsConstrained} can wrap our
approach into a practical workflow.

\paragraph{Overview.}

After giving key definitions in \Cref{sec:orgec395cf}, we present our
approach in \Cref{sec:orgc0ddf34}.  In the
\Crefrange{sec:org6038fb0}{sec:org4de5f97}, we explain the modelling,
in \Cref{sec:org781ed94} the treatment of collectives, in
\Cref{sec:org4d60f0a} the constrained POMDP synthesis problem,
and, in the \Cref{sec:strategy-import,sec:verification}, the
extraction and verification of a strategy.  We evaluate our approach
in \Cref{sec:orgc9fe360}, discuss issues we encountered during
modelling and synthesis in \Cref{sec:discussion}, and add concluding
remarks in \Cref{sec:conclusion}.

\section{Preliminaries}
\label{sec:orgec395cf}

Stochastic modelling is about describing uncertain real-world
behaviour in terms of states and probabilistic actions producing
transitions between these states.  For stochastic reasoning (i.e.,\xspace
drawing conclusions about such behaviour), we use \emph{probabilistic
  model checking}.  This section introduces the \emph{stochastic
  models}, \emph{temporal logic}, and \emph{tools} we employ 
for synthesis and verification.

\paragraph{Partially Observable Markov Decision Processes.}
\label{sec:orgf60ab5f}

Let $\text{\emph{Dist}}(X)$ denote the set of discrete probability
distributions over a set~$X$, and $\mathbb{R}_{\geq 0}$ be the non-negative
real numbers.  Then, a POMDP~\cite{Norman2017} is given by

\begin{definition}
  \label{def:pomdp}
  A POMDP is a tuple
  $M = (S, \overline{s}, A, P, R, \mathcal{O}, \mathit{obs})$, where
  \begin{itemize}
  \item $S$ is a set of states with $\overline{s} \in S$ being the
    initial state,
  \item $A$ is a set of actions (or action labels),
  \item $P\colon S \times A \to \mathit{Dist}(S)$ is a (partial)
    probabilistic transition function,
  \item $R = (R_S,R_A)$ is a structure defining state and action rewards 
    $R_S\colon S \to \mathbb{R}_{\geq 0}$ and
    $R_A\colon S \times A \to \mathbb{R}_{\geq 0}$,
  \item $\mathcal{O}$ is a finite set of observations, and
  \item $\text{\emph{obs}}: S \to \mathcal{O}$ is a labelling of
    states with observations.
  \end{itemize}
\end{definition}

Moreover, $A(s) = \{a \in A \mid P(s, a)~\text{is defined}\}$
describes the actions \emph{available} in $s$.  A \emph{path} in $M$
is defined as a finite or infinite sequence
$\pi = s_0 \stackrel{a_0}{\rightarrow} s_1 \stackrel{a_1}{\rightarrow}
\ldots$ where $s_i \in S$, $a_i \in A(s_i)$, and
$P(s_i,a_i)(s_{i+1})>0$ for all $i \in \mathbb{N}$.  Let
$\mathit{last}(\pi)$ be the last state of $\pi$.  $\mathit{FPaths}_M$
and $\mathit{IPaths}_M$ denote all finite and infinite paths of $M$
starting at state~$\overline{s}$.
Non-determinism in~$M$ is resolved through a \emph{strategy} according
to

\begin{definition}[POMDP Strategy]
  \label{def:pomdp-strategy}
  A strategy for a POMDP $M$ is a map $\sigma:
  \text{\emph{FPaths}}_M\rightarrow \text{\emph{Dist}}(A)$, where 
  \begin{itemize}
  \item for any $\pi\in\mathit{FPaths}_M$, we have $\sigma(\pi)(a)>0$
    only if $a \in A(\mathit{last}(\pi))$, and
  \item for any path
    $\pi = s_0 \stackrel{a_0}{\rightarrow} s_1
    \stackrel{a_1}{\rightarrow} \ldots$ and
    $\pi' = s_0' \stackrel{a_0'}{\rightarrow} s_1'
    \stackrel{a_1'}{\rightarrow} \ldots$ satisfying
    $\text{\emph{obs}}(s_i)=\mathit{obs}(s_i')$ and $a_i = a_i'$ for
    all $i$, we have $\sigma(\pi) = \sigma(\pi')$.
  \end{itemize}
\end{definition}
We call $\sigma$ \emph{memoryless} if $\sigma$'s choices
only depend on the most recent state ($\mathit{last}(\pi)$), and
\emph{deterministic} if $\sigma$ always selects an action with
probability~1.  Below, we consider memoryless deterministic
strategies.

\paragraph{Probabilistic Linear Temporal Logic (PLTL).}
\label{sec:org2f6eb21}

A POMDP $M$ describes a stochastic process, such that every
possible execution of that process corresponds to a \emph{path}
through $M$'s transition graph.  To draw qualitative conclusions
about~$M$ and its associated strategies, we express properties in
linear temporal logic~(LTL).  An LTL formula $\phi$ over atomic propositions
$\mathit{AP}$ follows the grammar
\begin{align}
  \label{grammar:pltl}
  \phi
  &::=
    \mathit{ap}
    \mid \neg\phi
    \mid \phi\land\phi
    \mid \tlop{X}\phi
    \mid \phi\mathrel{\tlop{U}}\phi
\end{align}
where $\mathit{ap}\in\mathit{AP}$.  In LTL, we make
statements about $M$'s path structure and specify admissible
sets of paths.  Informally, $\tlop{X}\phi$ describes that $\phi$ holds
in the next state of a given path, and $\phi\mathrel{\tlop{U}}\psi$ describes
that $\phi$ holds until $\psi$ occurs, or globally, if $\psi$ never
occurs.  We allow the abbreviations
$\tlop{F}\phi\equiv\textsc{t}\mathrel{\tlop{U}}\phi$, describing that $\phi$
holds at some point on a path, and
$\tlop{G}\phi\equiv\neg\tlop{F}\neg\phi$, describing that
$\phi$ applies to the entire path.

To draw quantitative conclusions about~$M$ (or query
probabilities and rewards), we use probabilistic LTL~(PLTL), whose formulas $\phi$
are formed along~\eqref{grammar:pltl} and by the two operators
$\mathop{\tlop{P}}$ and $\mathop{\tlop{R}}$:
\begin{itemize}
\item $\mathop{\tlop{P}}_{[\min\mid\max]\sim p\mid=?}[\psi]$ describes that the
  [minimum|maximum] probability of $\psi$ (under all possible
  POMDP strategies) being valid is~$\sim p$, and
\item $\mathop{\tlop{R}}^R_{[\min\mid\max]\sim r\mid=?}[\psi]$ expresses that the
  [minimum|maximum] expected reward~$R$ associated with~$\psi$ (under all
  possible POMDP strategies) meets the bound~$\sim r$,
\end{itemize}
where $\psi$ is an LTL formula, $\sim\,\in\{<,\leq,=,\geq,>\}$,
and $=?$ is used for queries.
Given a timer~$t$ in $M$ and that all actions in
$A$ increment $t$ by 1, we allow
$\tlop{F}^{\sim T}\psi\equiv\tlop{F}
(t\sim T\land\psi)$.
In LTL, $M\models\phi$ expresses that all paths of $M$
from~$\overline{s}$ are permitted by $\phi$, and, in PLTL, that a
probability measure over $M$'s paths satisfies~$\phi$.  For
convenience, we use $\phi$ to refer to both, the expression and the
region in $S$ where it evaluates to true.  LTL and
PLTL's semantics are explained in detail in, for example,
\cite[p.~231 and Sec.~6.2]{Baier2008-PrinciplesModelChecking}.

\paragraph{The \textsc{Prism}\xspace Model Checker}
\label{sec:orgd9d049a}

can check $M\models\phi$ using exact and approximate
algorithms~\cite{Parker2024-PRISMModelChecker}.  It supports a variety
of stochastic models and logics and has its own languages for modelling
and for specifying properties.  In \textsc{Prism}\xspace, a reward
structure~$R$ is used as a parameter
in~$\mathop{\tlop{R}}^R[\cdot]$.  For POMDPs, \textsc{Prism}\xspace can synthesise
strategies in the form of \Cref{def:pomdp-strategy}.

A \textsc{Prism}\xspace \emph{model} consists of a series of \emph{modules}, each
defining a fraction of a state~$s$ using its own variables.
\emph{Modules} can synchronise their transitions by sharing labels
from $A$, such that transitions in several modules using the same
label can only switch in a state $s\in S$ if each transition is
enabled in~$s$.  An example module with state variable $x$ and command
\textcolor{red}{\texttt{increase}} is given in
\Cref{lst:prism-example}.

{\begin{lstlisting}%
  [language=prism,
  caption={Example of a module in \textsc{Prism}\xspace's guarded command language},
  label=lst:prism-example]
module Example
  x : [0..2] init 0; // state variable $x$ with range $\{0, 1, 2\}$ and initial value $0$
  [increase] x<2 -> (x`=x+1); // command $\mathtt{increase}$ increasing $x$ by $1$ if $x<2$
endmodule 
\end{lstlisting}}

\paragraph{Fixed-Grid Approximation in \textsc{Prism}\xspace.}
\label{sec:orgae569ee}

When analysing a POMDP $M$, \textsc{Prism}\xspace computes an approximate
(finite-state) belief-MDP $B(M)$ \cite{Norman2017}, each
\emph{belief} being a probability distribution over the partially
observable states (e.g.,\xspace the possible room contamination).  The size of
$B(M)$'s state space, called \emph{belief space} \cite{fes}, is
exponential in \textsc{Prism}\xspace's grid-resolution parameter $g$
controlling the approximation of the upper and lower bounds to be
determined for $\mathop{\tlop{P}}|\mathop{\tlop{R}}^{\min|\max}$-properties using
$B(M)$.

\section{Developing Controllers by Example of the Cleaning Scenario}
\label{sec:orgc0ddf34}

\begin{figure}
  \centering
  \resizebox{\linewidth}{!}{
\includegraphics[width=\linewidth]{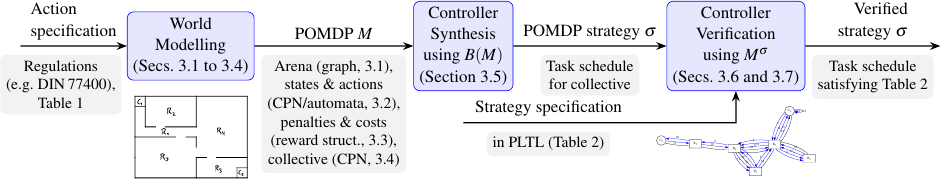}
 }
  \caption{Overview of the proposed synthesis approach for robot collectives}
  \label{fig:approach}
\end{figure}

In this section, we first state our synthesis problem and then
describe our approach to strategy synthesis and verification as
illustrated in \Cref{fig:approach}.

\paragraph{Problem Statement.}

Our aim to synthesise a collective controller is specified in the
LTL property
\begin{align}
  \label{frm:cond-bound-recur}
  \underbrace{
  \tlop{G}\big(
  \overbrace{
  (\omega\rightarrow\tlop{F}^{\leq T}(\omega\land\phi_{\mathsf{r}}))
  }^{\text{\color{blue} reach task goal}}
  \land
  \overbrace{
  \tlop{G}^{\leq T}\phi_{\mathsf{s}}
  }^{\text{\color{blue} keep safe}}
  \big)}_{\text{periodically achieve task safely (implied by our approach)}},
\end{align}
where $\omega$ is the \emph{recurrence area} (including
$\overline{s}$ and acting as a task invariant), $\phi_{\mathsf{r}}$ specifies an
\emph{invariant-narrowing condition},\footnote{$\phi_{\mathsf{r}}$ only has
  methodological relevance.  It could, for example, be used to develop
  an increasingly strong invariant $\omega$.}  $\phi_{\mathsf{s}}$
specifies \emph{task safety}, and $T>1$ is the
\emph{recurrence interval} (an upper cycle-time bound).
Among the controllers satisfying~\eqref{frm:cond-bound-recur}, we look
for an optimal (e.g.,\xspace one minimising energy consumption) and robust
(e.g.,\xspace under partial observability of stochastic room contamination)
one.

\paragraph{Overview of Controller Development.}

First, a spatio-temporal abstraction of the cleaning scenario is
modelled using a coloured Petri net~(CPN) for coordination modelling
and finite automata for describing robot-local behaviour.  These
aspects are translated into a reward-enhanced POMDP~$M$
(\Cref{sec:org6038fb0,sec:org052bad0,sec:org4de5f97}) in support of
multiple robots (\Cref{sec:org781ed94}), which uses probabilistic
actions to reduce the state count of a hypothetical detailed model.
Then, a strategy $\sigma$ is synthesised for $M$
(\Cref{sec:org4d60f0a}), which is used to derive a
\emph{deterministic}, \emph{non-probabilistic}, and
\emph{integer-valued} model~$M^\sigma$ (\Cref{sec:strategy-import}).
Finally, $M^\sigma$, representing the high-level controller, is verified
(\Cref{sec:verification}) against strategy requirements that, due to
current limitations in the formalisms and tools, cannot be checked
directly during synthesis.

\subsection{Spatio-temporal Abstraction}
\label{sec:org6038fb0}

\paragraph{State Space.}

For the implementation of the problem (e.g.,\xspace cleaning task), it is
important to keep the number of states as small as possible.
Therefore, large parts of the initial problem are abstracted.

\begin{figure}
  \centering
  \includegraphics[trim=0cm 1.5cm 0cm 1cm,height=2.5cm]{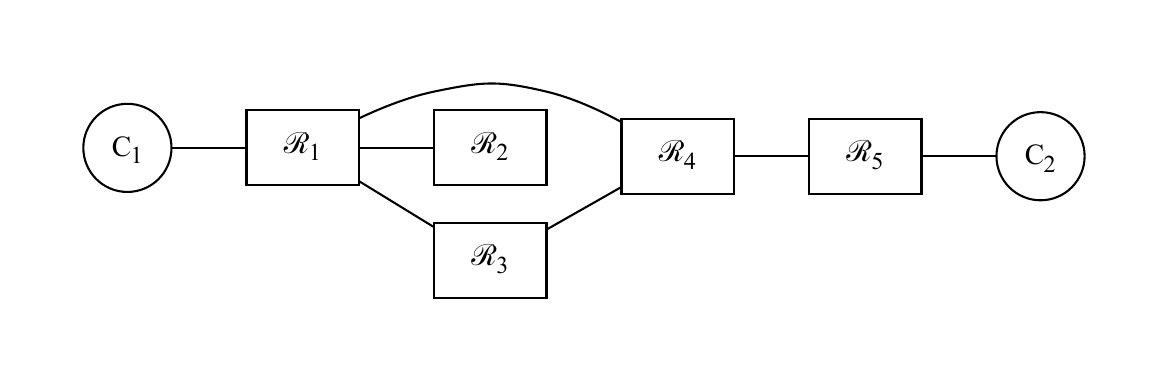}
  \caption{Example of a room plan graph\label{room-graph}}
\end{figure}

Instead of a complete room plan with area assignment, the abstracted
\emph{environment} uses a graph that only contains the different rooms
and charging stations.  An example of such a graph for the room plan
in \Cref{room-layout} can be seen in \Cref{room-graph}.  A pointer
$B_i.x$, $i\in 0..k$, to a room or charging station in this graph is
used to keep track of the position of robot $B_i$. The behaviour of
the charging state $B_i.c$ of the robot's battery is described by a
number of discrete charging levels and charge and discharge rates.

The total \emph{contamination} is represented by a counter $\mathcal{R}_j.d$,
$j\in 0..m$, whose maximum value is the contamination threshold
$\mathcal{R}_j.threshold$. Since we want to avoid reaching a state with the
contamination at this threshold, it is not necessary to model
contamination beyond $\mathcal{R}_j.threshold$.

\paragraph{Actions and High-level Behaviour.}

Discrete values are used to model \emph{time} as well, where the
action \prismid{at}, as specified in \Cref{tab:action-description} and
described below, is performed at each discrete time step.

\begin{table}
  \centering
  \small
  \caption{Requirements of the cleaning scenario defining the
    \prismid{at} action implemented in three modules}
  \label{tab:action-description}
  \begin{tabularx}{\linewidth}{cXl}
    \toprule
    \textbf{Id.}
    & \textbf{Action Specification} (Behavioural Requirement)
    & \textbf{Impl.~in Module}
    \\\midrule
    Ba & The robot either stays in the room it is currently in or moves to
    another room, whereby there must be an edge in the room plan graph
    between the current and next room. 
    & \prismid{cleaner}
    \\Bb & If the robot is on a charging station, its battery charge
    level is increased by the charging rate of the robot. 
    \\Bc & If the robot is \emph{not} on a charging station, its battery
    charge level is reduced by the robot's discharge rate. 
    \\\midrule
    Da & The total contamination of each room increases by its
    contamination rate if there is no robot in that room. 
    &\prismid{contamination}
    \\
    Db & If the robot is in a room, the total contamination of that
    room is reset.
    \\\midrule
    Ta & The time counter is incremented by one.
    & \prismid{time} 
    \\\bottomrule
  \end{tabularx}
\end{table}

\begin{figure}
  \centering
  \subcaptionbox{Coordination of $k\leq n$ cleaners\label{fig:bot-coordinator}}[.45\linewidth]%
  {\includegraphics[scale=1]{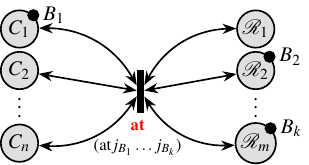}}
  \subcaptionbox{Control of cleaner $B_i$\label{fig:bot-controller}}[.45\linewidth]%
  {\includegraphics[scale=1]{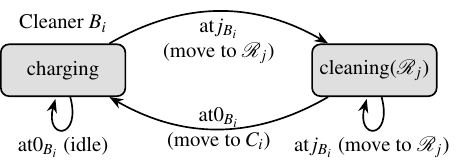}}
  \caption{Cleaner coordination (a) as a CPN and local
    control (b) as a finite automaton}
  \label{fig:cleaner-logic}
\end{figure}

The CPN in \Cref{fig:bot-coordinator} provides a
high-level description of the moves of the collective~$B$
across charging stations~$C$ and
rooms~$\mathcal{R}$~(\Cref{room-graph}).  The abstract \prismid{at} action
(black bar) expands to a range of concrete POMDP actions
\prismid{at$j_{B_1}\dots j_{B_k}$} with $j_{B_i}\in 0..m$.  Whenever
\prismid{at} is taken, any number of tokens~(black dots, representing
robots) on the places~(grey circles, representing rooms and charging
stations) can flow simultaneously, such that $B_i$ can move from one
place via \prismid{at} to an adjacent empty~(possibly same)
place.\footnote{We assume for any initial state $\overline{s}\in S$ that
  no more than one robot is at a particular place.}
\Cref{fig:bot-controller} outlines the control of a particular
robot~$B_i$.  When composed~(in parallel), as formalised in $M$
by implicit constraints on the~$j$-indices, simultaneous moves of
several robots into a single place and jumps to non-adjacent places
are prohibited by the coordination constraint in
\Cref{fig:bot-coordinator}.\footnote{This construction reduces $S$ and
  $P$ of $M$ in comparison with using alphabetised synchronous
  composition.}

\subsection{Quantitative and Stochastic Abstraction}
\label{sec:org052bad0}

For the sake of simplicity, this section and the following will focus
on a reduced problem with only one robot.  The case of multiple robots
will be reintroduced in \Cref{sec:org781ed94}.

As already mentioned, the \textsc{Prism}\xspace-encoding of $M$ is divided into
three modules operating on three independent fragments of the state
space~$S$: The state of the robots, the room contamination, and time.
\begin{description}
\item[The \prismid{cleaner} module] describes the behaviour of
  cleaning robot $B_1$.  An integer is used to model the robot
  position $B_1.x$.  For this, each room and charging station is
  mapped to an integer bijectively.  For example, the room graph in
  \Cref{room-graph} can be described by the following relation:
  \(C_0\rightarrow0, \mathcal{R}_1\rightarrow1, \mathcal{R}_2\rightarrow3,
  \mathcal{R}_3\rightarrow4, \mathcal{R}_4\rightarrow5, \mathcal{R}_5\rightarrow6,
  C_2\rightarrow7\).  The battery status $B_1.c$ is also described as
  an integer whose upper bound is the maximum charge $B_1.maxcharge$
  of $B_1$'s battery, see \Cref{lst:cleaner-states}.

{\begin{lstlisting}[
    label=lst:cleaner-states,
    caption={A model fragment of the \emph{cleaners} module highlighting its states and actions \label{cleaner-example1}}]
module cleaners
  x : [0..$m-1$] init $\mathit{B_1.start}$; // position of cleaner $B_1$
  c : [0..$\mathit{B_1.maxcharge}$] init $\mathit{B_1.{\omega}chgthres}$; // battery status of $B_1$
  [at0] (x=0|x=1)
         -> (x'=0) & (c=min(c+$\mathit{B_1.chargerate}$, $\mathit{B_1.maxcharge}$))(*\label{lst:increase}*); // charge $B_1$
  [at1] (x=0|x=1|x=2|x=3|x=4)
         -> (x'=1) & (c=max(c-$\mathit{B_1.dischargerate}$, 0))(*\label{lst:decrease}*); // move to $\mathcal{R}_1$
  ...
endmodule
\end{lstlisting}}

For each charging station and each room there is a transition \mytexttt{atN} (where \mytexttt{N} is the integer assigned to the room), which models entering or staying in this room.
The precondition for this transition is that the robot must already be in that room or a neighbouring room.
If a robot enters or stays at a charging station, the charge increases by the charging rate; if a robot is in a room, the charge decreases by the discharge rate, see lines \ref{lst:increase} and \ref{lst:decrease} respectively.

\item[The \textcolor{red}{\texttt{contamination}} module] describes
  the contamination status of $\mathcal{R}$.  To reduce the number of states,
  contamination is modelled by booleans---the contamination flags
  $\mathcal{R}_j.d$, $j\in 1..m$---rather than integers.  The probability
  $\mathcal{R}_j.pr$ of $\mathcal{R}_j.d$ getting true is used to model the state in
  which the contamination of the corresponding room has reached its
  threshold~$\mathcal{R}_j.threshold$.
  If there is no robot in $\mathcal{R}_j$, we set $\mathcal{R}_j.d =$
  \mytexttt{true} with probability $\mathcal{R}_j.pr$ inversely proportional to
  the contamination threshold $\mathcal{R}_j.threshold$.  If a robot visits or
  stays in $\mathcal{R}_j$ then $\mathcal{R}_j.d$ is set to \mytexttt{false}, see
  \Cref{lst:dirtmodule}.
\end{description}

{\begin{lstlisting}[language=prism,
  label=lst:dirtmodule,
  caption={A fragment of the \prismid{contamination} module (e.g.\ $\mathcal{R}_i.pr=0.05$
    for $i\in1..m$) \label{dirt-example1}}]
module contamination // sequential stochastic contamination
  $\mathcal{R}_1.d$ : boolean init false; $\mathcal{R}_2.d$ : boolean init false;
  ...
  [at0] true -> $1-(\sum_{i\in1..m} \mathcal{R}_i.pr)$:  true
     + $\mathcal{R}_1.pr$: ($\mathcal{R}_1.d$'=true) + $\mathcal{R}_2.pr$: ($\mathcal{R}_2.d$'=true) + ...; // prob. contam. all while charg.
  [at1] true -> $1-(\sum_{i\in1..m\setminus 1} \mathcal{R}_i.pr)$: ($\mathcal{R}_1.d$'=false)
     + $\mathcal{R}_2.pr$: ($\mathcal{R}_1.d$'=false)&($\mathcal{R}_2.d$'=true) + ...; // clean $\mathcal{R}_1$ and prob. cont. other rooms
  [at2] true -> $1-(\sum_{i\in1..m\setminus 2} \mathcal{R}_i.pr)$: ($\mathcal{R}_2.d$'=false)
     + $\mathcal{R}_1.pr$: ($\mathcal{R}_2.d$'=false)&($\mathcal{R}_1.d$'=true) + ...; // clean $\mathcal{R}_2$ and prob. cont. other rooms
  ...
endmodule
\end{lstlisting}}

\begin{description}
\item[The \prismid{time} module] describes the progression of time and
  manages the switching to the error and final state (the model
  handles both the same).  \prismid{time} also restricts the
  \mytexttt{atN} transitions so that they can only be used as long as
  the model is not in the error or final state.  This is
  possible because a transition can only trigger if it can trigger in
  each module. Therefore, precondition in the \prismid{time} module
  can prevent a transition from triggering even though it is marked
  with \mytexttt{true} in the \prismid{cleaner} and \prismid{contamination}
  modules.  The time is increased by one unit with each transition
  until it reaches~$T$. At this point (due to the definition
  of \mytexttt{error\_or\_final}), only the \mytexttt{fin} transition
  can switch and the model ends in a loop, see
  \Cref{lst:time-structure}.
\end{description}
{\begin{lstlisting}[
  label=lst:time-structure,
  caption={A fragment of the \prismid{time} module \label{time-example1}}]
formula error_or_final = (c=0|!(t<T));
module time
  t : [0..T] init 0;
  [at0] !error_or_final -> (t'=min(t+1,T)); // charge
  [at1] !error_or_final -> (t'=min(t+1,T)); // move to $\mathcal{R}_1$
  [fin] error_or_final  -> (t'=min(t+1,T)); // finish cycle
endmodule
\end{lstlisting}}

\subsection{Choice of the Reward Function} 
\label{sec:org4de5f97}

Four requirements, a valid strategy must satisfy, can be derived from
Formula~\eqref{frm:cond-bound-recur} and \Cref{tab:action-description}:

\begin{itemize}
\item[FR] At time $T$, all robots must be back in their
  initial location, so that the plan can be repeated.
\item[$\omega$C] At time $T$, the battery of a robot must
  not be lower than its threshold charge level.
\item[BC] The battery of a robot must never be empty.
\item[CT] The total contamination of any room must never exceed its
  contamination threshold.
\end{itemize}

When just focusing MDP verification rather than synthesis, it
would be sufficient to describe these as PLTL constraints.
However, PLTL constraints cannot be used as queries for
synthesising \emph{strategies}, since generating strategies through
\textsc{Prism}\xspace requires each path to be able to fulfil all constraints
eventually.
Using constraints that are violated on some of $M$'s paths~(e.g.,\xspace
if we require BC, any path leading to an empty battery eventually
violates BC) will prevent \textsc{Prism}\xspace from finding a reward-optimal
strategy.  This is a specific known limitation of the used formalism.
Even though we cannot use PLTL constraints to encode all our
requirements, the reward structure~$R$ can be used to prioritise 
selecting \emph{strategies} that fulfil these requirements.
The encoding of our requirements in $R$ can be achieved by penalising
states that do not fulfil some or all of the requirements, see
\Cref{lst:penalties}.

{\begin{lstlisting}[
  label=lst:penalties,
  caption={An example of the reward structure for the \emph{state penalities}}]
const a_lot = 10000000;
const a_bit = 10000;
rewards "penalties"
  c=0:                               a_lot; // BC: battery empty
  t=T & (x!=$\mathit{B_1.start}$|c<$B_1.\omega$chgthres):  a_lot; // FR: robot not at initial loc. at time $T$
  $\mathcal{R}_1.d$=true:                           a_bit; // Room 1's contamination flag is set
  $\mathcal{R}_2.d$=true:                           a_bit; // Room 2's contamination flag is set
  ...
endrewards
\end{lstlisting}}

We previously found it ineffective to penalise the contamination flags
the same as the constraints FR, $\omega$C, and BC.  Whereas the latter can
be determined from $M$'s state, contamination flags only carry
the probability of a requirement being violated.  Hence, we apply
lower penalties to the contamination flags.
Further reward structures are used to model optimisation goals, such
as energy consumption, see \Cref{lst:energycons}.  However, the
penalty for constraints is chosen such that it is not possible
to offset the penalty of an invalid state by the reduced penalty for a
less energy-consuming strategy.

\begin{figure}[h]
  \centering
  \setcaptiontype{lstlisting}
  \begin{minipage}[t]{0.45\textwidth}
    {\begin{lstlisting}
rewards "energy consumption"
  t < T: $\mathit{B_1.maxcharge}$ - c; 
endrewards
\end{lstlisting}}
    \subcaption{A fragment of the optimisation \emph{rewards}}
    \label{lst:energycons}
  \end{minipage}
  \hfill
  \begin{minipage}[t]{0.45\textwidth}
    {\begin{lstlisting}
const a_lot = 10000000;
rewards "utilisation"
  x=2 & t>=8  & t<10: a_lot;
  x=2 & t>=12 & t<14: a_lot; ...
endrewards
\end{lstlisting}}
    \subcaption{A fragment of room utilisation
      \emph{rewards}}
    \label{lst:roomutil}
  \end{minipage}
  \caption{Fragments of the reward structure used for the cleaning
    scenario}
  \label{fig:rewardstructure}
\end{figure}

\paragraph{Room Utilisation.}
\label{sec:org5ea3598}

As with the other requirements, a room utilisation profile to be
respected by the cleaners~(UT) can be softly specified using
additional reward functions as shown in \Cref{lst:roomutil}.  However,
UT as a safety property will later also be specified in PLTL and
checked of the synthesised strategy.

\subsection{Cooperation between Multiple Robots}
\label{sec:org781ed94}

\begin{wrapfigure}[6]{r}{7cm}
  \vspace{-3em}
  {\begin{lstlisting}
x$_1$ : [0..$m$] init $B_1.start$;
x$_2$ : [0..$m$] init $B_2.start$;
...
c$_1$ : [0..$B_1.maxcharge$] init $B_1.maxcharge$;
c$_2$ : [0..$B_2.maxcharge$] init $B_2.maxcharge$;
...
\end{lstlisting}}
  \vspace{-1em}
  \caption{The structure of the \prismid{cleaner} state}
  \label{lst:robotstatus}
\end{wrapfigure}

To keep $M$ simple, robots are not modelled as separate modules,
but the state of the \prismid{cleaner} module is extended to include
the positions of all robots (see \Cref{fig:bot-coordinator}).  This
simplification excludes all transitions from the model that would lead
to conflicts in robot behaviour (e.g.,\xspace the case where several robots
clean the same room at the same time).  Additionally, each robot has
its own battery charge, see Listing~\ref{lst:robotstatus}.

The \mytexttt{atN} actions for a single robot are now extended to
\mytexttt{atN\_N\_...} actions, which then model the simultaneous
movement of $k$ robots, as illustrated in \Cref{fig:bot-coordinator}
and implemented in \Cref{lst:robottrans}.

{\begin{lstlisting}[
  label=lst:robottrans,
  caption={Two examples of \mytexttt{atN\_N} actions}]
[at0_1] !error & (x$_1$=0|x$_1$=1) & (x$_2$=0|x$_2$=1|x$_2$=2|x$_2$=3|x$_2$=4) // charge $B_1$ and move $B_2$ to $\mathcal{R}_1$
         -> (x$_1$'=0) & (x$_2$'=1)
          & (c$_1$'=min(c$_1$+$B_1.chargerate$,$B_1.maxcharge$))
          & (c$_2$'=max(c$_2$-$B_2.dischargerate$,0));
[at1_2] !error & (x$_1$=0|x$_1$=1|x$_1$=2|x$_1$=3|x$_1$=4) & (x$_2$=1|x$_2$=2) // move $B_1$ to $\mathcal{R}_1$ and $B_2$ to $\mathcal{R}_2$
         -> (x$_1$'=1) & (x$_2$'=2)
          & (c$_1$'=max(c$_1$-$B_1.dischargerate$,0))
          & (c$_2$'=max(c$_2$-$B_2.dischargerate$,0));
...
\end{lstlisting}}

This solution increases the number of states per robot considerably,
but the complexity of $M$ is still within a practically
verifiable range.
Additionally, the reward structures that depend on the position and
charge of a robot are adapted to include the position of all robots,
as can be seen in \Cref{lst:robottransreward}. 

{\begin{lstlisting}[
  label=lst:robottransreward,
  multicols=2,
  caption={Reward structure for a collective}]
rewards "penalties"
  c$_1$=0: a_lot;
  c$_2$=0: a_lot;
  ...
  t=T & (x$_1$!=$B_1.start$|c$_1$<$B_1.{\omega}chgthres$): a_lot;
  t=T & (x$_2$!=$B_2.start$|c$_2$<$B_2.{\omega}chgthres$): a_lot;
  ...
endrewards
  
rewards "utilisation"
  x$_1$=2 & t>=8  & t<10: a_lot;
  x$_2$=2 & t>=8  & t<10: a_lot;
  x$_1$=2 & t>=12 & t<14: a_lot;
  x$_2$=2 & t>=12 & t<14: a_lot;
  ...
endrewards
\end{lstlisting}}

\subsection{Synthesising Strategies (under Uncertainty) for the Cleaning Scenario}
\label{sec:org4d60f0a}

\textsc{Prism}\xspace's POMDP strategy synthesis works under certain
limitations.  As indicated in \Cref{sec:org4de5f97}, it is not
possible to use $\mathop{\tlop{R}}^R_{\min=?}[\psi]$ for
synthesis if $\mathop{\tlop{P}}_{\min=?}[\psi]<1$, that is, if $M$ contains
$\psi$-violating paths under some strategy $\sigma$.  Hence, we
choose a $\psi$ that defines a state that all paths converge at, and
synthesise a strategy that minimises the total reward (since we
model $R$ using penalties) up to that point.  A
commonality of all paths is the flow of time, so the reachability
reward-based synthesis query
\begin{align}
  \label{synthesis-query}
  \textstyle
  \sum_{R\in\{\text{penalties, energy consumption, utilisation}\}}
  \mathop{\tlop{R}}^R_{\min=?} [\tlop{F}\,t=T]
\end{align}
uses a target state where time $t$ is equal to some maximum time
$T$.

\begin{wrapfigure}[6]{r}{3cm}
{\begin{lstlisting}
observables 
  t,
  x$_1$, x$_2$, ...,
  c$_1$, c$_2$, ...
endobservables
\end{lstlisting}}
\end{wrapfigure}

Additionally, a mapping $obs$ needs to be specified, which defines the
observations of $M$ that $\sigma$ can use to make choices.  In
this case, $\sigma$ cannot use the contamination flags to make its
choices.  If $\sigma$ could consider the contamination flag, it
would not need to account for the accumulative probability;
$\sigma$ could just check if a contamination flag is true and act
accordingly.  We can hide the contamination flags from $\sigma$ by
defining $obs$ to just include the position and charge of the robot
and the time, see the listing on the right.

\textsc{Prism}\xspace allows the explicit generation of \emph{deterministic}
strategies.  Such strategies are useful in this case, since, 
except for the final state, the model has no loops (i.e.,\xspace time is
always advancing).  
Because $\sigma$ is deterministic, it can be thought of as a
list of actions for each time step of the cleaning schedule, where the
transition of each step of the strategy denotes the action of the
robot at that time step within period $T$.  Note that the
observable environmental part of $M$ is deterministic such that
after applying $\sigma$, $M^\sigma$ has exactly one path, hence,
$\sigma$ only depends on time~$t$.

\subsection{Creating an Induced Model from the Strategy}
\label{sec:strategy-import}

It is possible to create a cleaning schedule from the synthesised
strategy $\sigma$.  However, it is not yet possible to verify
$\sigma$ regarding the constraints listed in \Cref{sec:org4de5f97}.
This is because, up to this point, contamination was modelled in
$M$ only as a probabilistic factor.  To verify the contamination
constraint CT, below, we include a non-probabilistic contamination
model in $M'$ using counters to represent contamination.

\paragraph{Modelling the Contamination Value.}

To verify that the contamination value (modelled as a boolean
sub-MDP of $M$) never actually reaches the thresholds
$\mathcal{R}_j.threshold$, $j\in 1..m$, it
is necessary to transform $M$ into an MDP
$M'$ that accounts for the actual values $\mathcal{R}_j.d$.  We
accomplish this in $M'$ by integer-valued contamination counters
$\mathcal{R}_j.d$ (\Cref{lst:dirt-new}) replacing the boolean variables
$\mathcal{R}_j.d$ in $M$ (\Cref{lst:dirtmodule}).

{\begin{lstlisting}[
  label=lst:dirt-new,
  caption={An example of the structure of the \emph{contamination} module using discrete contamination values\label{dirt-example2}}]
module contamination
  $\mathcal{R}_1.d$ : [0..$\mathcal{R}_1.threshold$] init 0;
  $\mathcal{R}_2.d$ : [0..$\mathcal{R}_2.threshold$] init 0;
  ...
  [at0_6] true -> ($\mathcal{R}_1.d$'=min($\mathcal{R}_1.d$+$\mathcal{R}_1.contaminationrate$,$\mathcal{R}_1.threshold$))
                & ($\mathcal{R}_2.d$'=min($\mathcal{R}_2.d$+$\mathcal{R}_2.contaminationrate$,$\mathcal{R}_2.threshold$)) & ...;
  [at0_1] true -> ($\mathcal{R}_1.d$'=0)
                & ($\mathcal{R}_2.d$'=min($\mathcal{R}_2.d$+$\mathcal{R}_2.contaminationrate$,$\mathcal{R}_2.threshold$)) & ...;
  [at0_2] true -> ($\mathcal{R}_1.d$'=min($\mathcal{R}_1.d$+$\mathcal{R}_1.contaminationrate$,$\mathcal{R}_1.threshold$))
                & ($\mathcal{R}_2.d$'=0) & ...;
  ...
endmodule
\end{lstlisting}}

\paragraph{Applying the Strategy.}

\begin{wrapfigure}[8]{r}{5.2cm}
{\begin{lstlisting}
module time
  ...
  [at0_1] !error_or_final
          & (t=8|t=10)
          -> (t'=min(t+1,T))
  ...
endmodule
\end{lstlisting}}
\end{wrapfigure}

Apart from using contamination counters, our model does no longer
contain probabilistic choices.  Concretely, each probabilistic choice
in $M$ (branching to each possible selection of fully
contaminated rooms, \Cref{lst:dirtmodule}) is replaced by an action
in~$M'$ performing a simultaneous update of all contamination
counters~(\Cref{lst:dirt-new}).
We can use the generated strategy $\sigma$ to derive the induced
deterministic model $M^\sigma$ from $M'$, which acts according
to the strategy.
This step is done by modifying the preconditions of the
\mytexttt{atN\_N\_...} actions of the \prismid{time} module to only be
able to trigger when that action is chosen at the same point in the
strategy.  If, for example, within
$\sigma$, the \mytexttt{at0\_1} action is only chosen in time steps
8 and 10, we modify the \prismid{time} module, see the listing on the
right.

\paragraph{Notes on the Relationship between $M$, $M'$, and
  $M^\sigma$.}

The state space of $M'$ is significantly larger than the one of
$M$ as the latter contains intermediate contamination $\mathcal{R}_j.d$
up to $\mathcal{R}_j.d=\mathcal{R}_j.\mathit{threshold}$.  The fact that $\mathcal{R}_j$ gets
contaminated in $M$ corresponds to all shortest sequences of
transitions with non-zero probability to a state where $\mathcal{R}_j.d=$ true.
The same fact in $M'$ corresponds to all sequences of transitions
leading to $\mathcal{R}_j.d=\mathcal{R}_j.\mathit{threshold}$.  Hence, the time module
in $M^\sigma$ is a refinement of the time module in $M$ and, due
to synchronisation (via action labels), the resetting of
contamination's (i.e.,\xspace the cleaning) in $M^\sigma$ is a refinement of
the corresponding resets in~$M$.

A property of $M'$ preserving quantitative strategy correctness,
that we left for future work, is to check whether the probabilities of
the contamination flags set in $M$ are greater than or equal to
the hypothetical probabilities of the corresponding counters in
$M'$ and, thus, $M^\sigma$, reaching their thresholds.  This
property, when true, expresses that the flags are a sound (i.e.,\xspace
conservative) quantitative abstraction of the counters.

\subsection{Strategy Verification via Verifying the Induced Model}
\label{sec:verification}

Now, we use~$M^\sigma$ to check \emph{recurrence} and \emph{safety}
from Formula~\eqref{frm:cond-bound-recur}, that is, 
$\omega\to\tlop{F}^{\leq T}\omega$ and
$\tlop{G}^{\leq T}\phi_{\mathsf{s}}$.\footnote{For the sake of
  simplicity of the example, we use $\phi_{\mathsf{r}}\equiv\top$ and can
  omit $\omega\to\tlop{F}^{\leq T}\phi_{\mathsf{r}}$.}
In particular, we check their decomposed translations into PLTL
requirements\footnote{All properties are expressed in quasi-LTL, that is,
  ACTL* allowing only one universal quantifier at the outermost
  level.} listed in \Cref{tab:strat-req}.  The recurrence
area~$\omega$ is encoded by the state propositions in FR,
$\omega R$, and $\omega C$, while safety~$\phi_{\mathsf{s}}$ is
encoded by the state propositions in BC, CT, and UT.  Note that the
upper bound~$T$ of the recurrence interval is met by all
paths in $M^\sigma$.  $\mathit{util}(\mathcal{R}_i)$ is the set of time slots
in which $\mathcal{R}_i$ is utilised.

For checking
$\omega\to\tlop{F}^{\leq T}\omega$, we
define $\omega$ to be a (not necessarily maximum) region in
$S$ from where $\sigma$ can be applied and $\omega$
is revisited after $T$ steps.
The requirements for $\sigma$ need to be true for every initial
state in $\omega$.  FR, $\omega R$, and $\omega C$ ensure
that applying $\sigma$ leads to a state within $\omega$.  To
specify $\omega$, we use thresholds for the battery charge of
robots ($B_i.c$ for every robot $B_i$) and the contamination of rooms
($\mathcal{R}_i.\omega contthres$ for every room $\mathcal{R}_i$).  $\omega$ then
characterises every state of $M^\sigma$ where the battery charge and
room contamination are within these thresholds and all robots are on
their starting position.  Recurrence can even be checked more easily for
$M^\sigma$ by selecting the worst state in $\omega$, which is the
state where every value lies exactly at the thresholds, and verifying
the requirements in \Cref{tab:strat-req} for this state.

\begin{table}
  \centering
  \small
  \caption{Requirements for validating the
    synthesised strategies (checks of $M^\sigma$ by \textsc{Prism}\xspace)} 
  \label{tab:strat-req}
  \begin{tabularx}{\linewidth}{c>{\hsize=.53\hsize}X>{\hsize=.47\hsize}X}
    \toprule
    \textbf{Id.}
    & \textbf{Strategy Specification} (Behavioural Requirement)
    & \textbf{\dots expressed in PLTL}
    \\\midrule
    FR
    & Cleaner $B_i$ Finally Returns to its starting position.
    & $\bigwedge_{i\in[1..k]}\mathop{\tlop{P}}_{\geq 1}[\tlop{F}^{=T}
    B_i.x=\mathit{B_i.start}]$
    \\
    $\omega R$
    & Cleaner $B_i$ has a final charge of at least $\mathit{B_i.\omega chgthres}$.
    & $\bigwedge_{i\in[1..k]}\mathop{\tlop{P}}_{\leq 0}[\tlop{F}^{=T}
     B_i.c<\mathit{B_i.\omega chgthres} ]$
    \\
    $\omega C$
    & Contamin.\ of $\mathcal{R}_i$ is finally less than $\mathit{\mathcal{R}_i.\omega contthres}$.
    & $\bigwedge_{i\in[1..m]}\mathop{\tlop{P}}_{\leq 0}[\tlop{F}^{=T}\mathcal{R}_i.d<\mathcal{R}_i.\omega contthres ]$
    \\\midrule
    BC
    & Battery charge of Cleaner $B_i$ is never 0.
    & $\bigwedge_{i\in[1..k]}\mathop{\tlop{P}}_{\leq0}[\tlop{F}\, B_i.c = 0 ]$
    \\
    CT
    & Contamination of $\mathcal{R}_i$ never exceeds $\mathcal{R}_i$'s threshold.
    & $\bigwedge_{i\in[1..m]}\mathop{\tlop{P}}_{\leq0}[\tlop{F}\, \mathit{\mathcal{R}_i.d} \geq \mathit{\mathcal{R}_i.threshold} ]$
    \\
    UT
    & Room $\mathcal{R}_i$ is not cleaned while occupied.
    &
    $\bigwedge_{i\in[1..m]}\bigwedge_{T\in\mathit{util}(\mathcal{R}_i)}\bigwedge_{j\in[1..k]}
    \mathop{\tlop{P}}_{\leq0}[\tlop{F}^{=T} B_j.x=\mathcal{R}_i ]$
    \\\bottomrule
  \end{tabularx}
\end{table}

\section{Experimental Evaluation}
\label{sec:orgc9fe360}

Our experimental evaluation addresses two research questions (RQs).

\begin{figure}[t]
  \centering
  \subcaptionbox{Strategy for one robot\label{strat1}}[\linewidth]
  {\includegraphics[height=5cm,trim=38 38 38 38,clip]{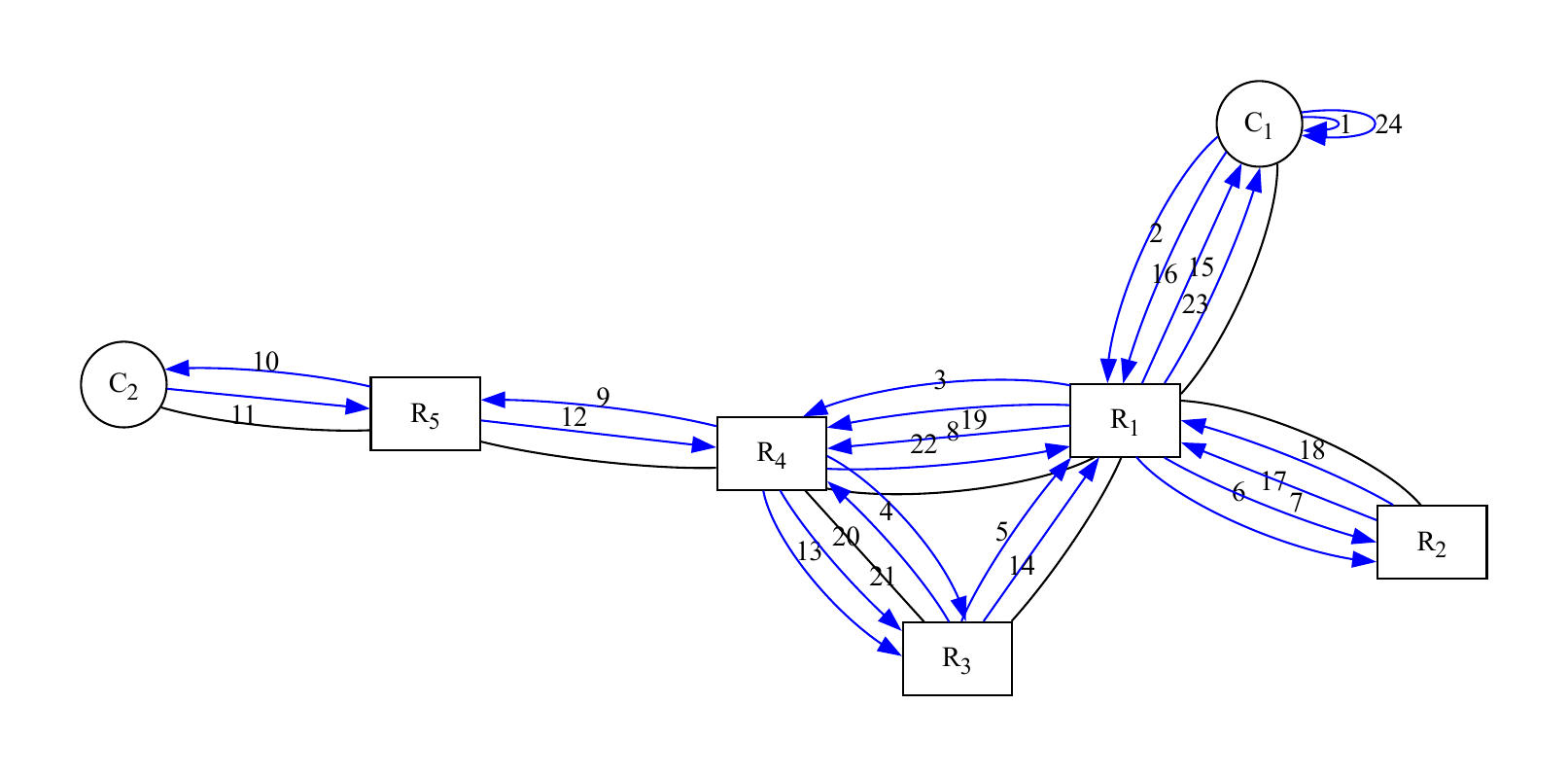}}
  \subcaptionbox{Strategy for two robots (blue and green directed arcs)\label{strat2}}
  {\includegraphics[height=5cm,trim=38 38 38 38,clip]{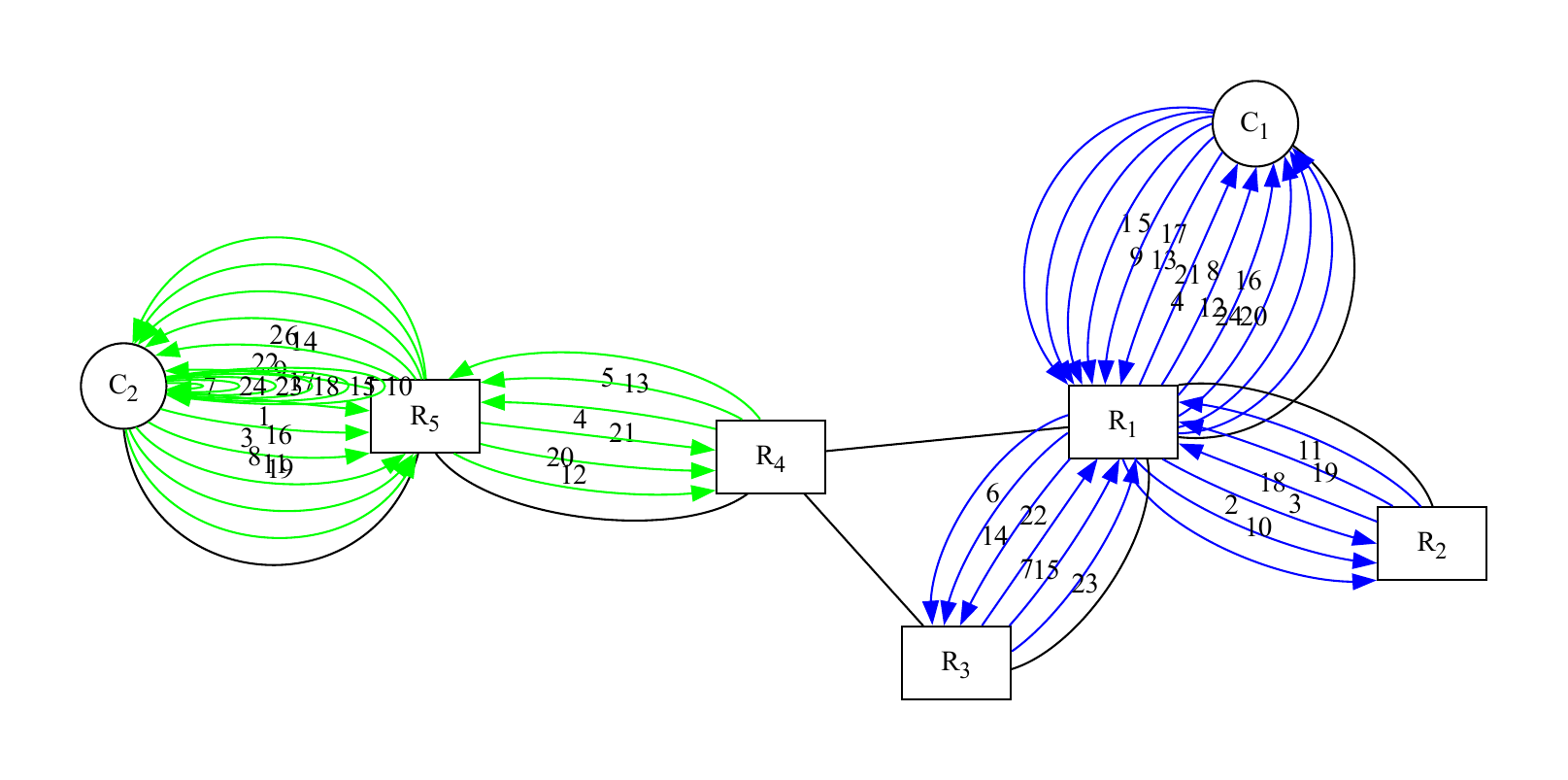}}
  \medskip
  
  \subcaptionbox{Execution of a 24h model cycle using the strategy in
    \Cref{strat2}; (sub-)cycles indicated in bold
    \label{fig:run}}{
    \footnotesize
    \begin{tabular}{c|cccccccccccccccccccccccc}
      \toprule
      t     &0 & 1& 2& 3& 4& 5& 6&   7& 8& 9&  10&11&12&13&14&15&\dots&24\\\midrule
      B$_1$ &\textbf{C$_1$}&$\mathcal{R}_1$&$\mathcal{R}_2$&$\mathcal{R}_1$&C$_1$&$\mathcal{R}_1$&$\mathcal{R}_3$&  $\mathcal{R}_1$&\textbf{C$_1$}&$\mathcal{R}_1$&  $\mathcal{R}_2$&$\mathcal{R}_1$&C$_1$&$\mathcal{R}_1$&$\mathcal{R}_3$&$\mathcal{R}_1$&\dots&\textbf{C$_1$}\\
      B$_2$ &\textbf{C$_2$}&$\mathcal{R}_1$&C$_2$&$\mathcal{R}_5$&$\mathcal{R}_4$&$\mathcal{R}_5$&C$_2$&idle&$\mathcal{R}_5$&C$_2$&idle&$\mathcal{R}_5$&$\mathcal{R}_4$&$\mathcal{R}_5$&C$_2$&idle&\dots&\textbf{C$_2$}
      \\\bottomrule
    \end{tabular}}
  \caption{Synthesised strategies.  Nodes represent charging stations
    and rooms; undirected arcs indicate room connections (doors);
    edge labels specify the execution order of particular \prismid{at}
    actions.} 
  \label{strategies}
\end{figure}

\subsection{RQ1: Can we synthesise reasonable strategies for multiple
  robots?}
\label{sec:rq1}

In the following, an instance of the example model with only one robot
is considered first. The contamination rate is the same for all rooms,
except that \(\mathcal{R}_5.d\) has a threshold value \(\mathcal{R}_5.threshold\) of 24
(based on a contamination rate of 1\,h$^{-1}$), which is twice as
high, whereas all other rooms have a threshold value of 12.  We
identified $\omega$ manually by examining the generated strategy.
Using the method described in \Cref{sec:orgc0ddf34}, a strategy was
generated that meets all the requirements.  This strategy is
visualised in \Cref{strat1}.  Each action is shown with a blue arrow,
at which the time step in which the action is to be executed is
annotated.
To develop a strategy for two robots, the battery charge was halved to
keep the model less complex. The corresponding strategy can be seen in
\Cref{strat2}.
This result turns out to be a \emph{partition-based patrolling
  strategy}, considered effective under random
disturbance~\cite[143]{Portugal2011-SurveyMultirobot}.

\subsection{RQ2: How do model parameters influence the synthesis of recurrent strategies?}
\label{sec:experiment-parameter}

\begin{figure}[t]%
\centering
\includegraphics[trim=0cm 0cm 0cm 0cm,clip,width=\textwidth]{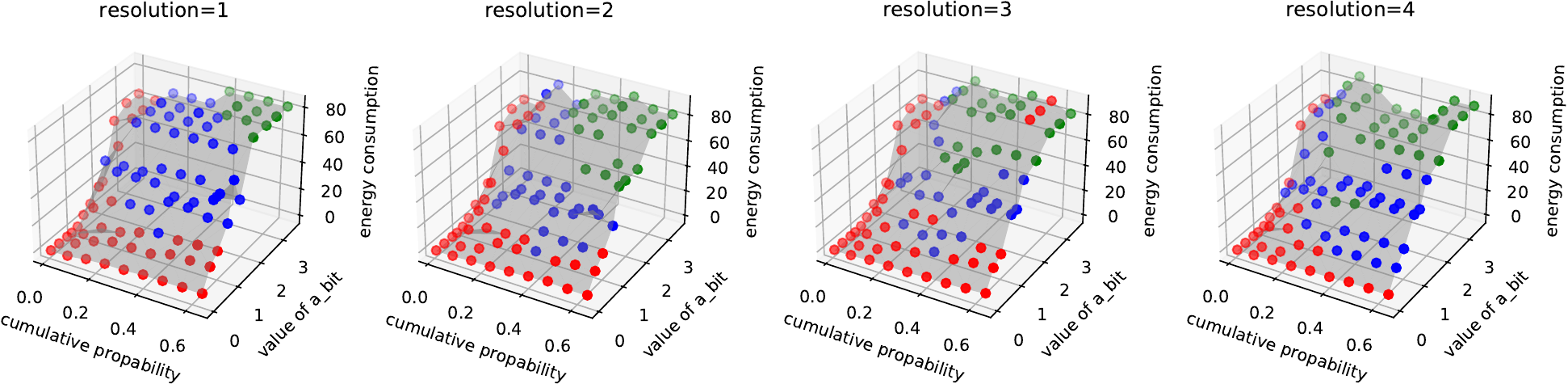}
\caption{Parameterised evaluation. Non-recurrent strategies are marked in blue, recurrent strategies in green.\label{runs} The scale for $a\_bit$ is logarithmic.}
\end{figure}

We evaluate the model using $\mathit{a\_bit}$,
$\mathit{\mathcal{R}_i.pr}$, and the fixed-grid resolution $g$ as parameters.
\Cref{runs} visualises the result using these parameters on a simplified model, which only contains one robot and omits room $\mathcal{R}_5$ and charger $C_2$. 
When building $\sigma$, we set $\mathit{\mathcal{R}_i.pr} =
\mathit{cumulative\_probability} / \vert
\mathcal{R}\vert$. 
The generated strategies were verified by iteratively assuming
$\omega$-thresholds (\Cref{tab:strat-req}) from a set chosen appropriately.

\Cref{runs} contains four plots for resolutions $g=1..4$.
Correct non-recurrent strategies are represented by a \textcolor{blue}{blue} dot, correct recurrent strategies by a \textcolor{green}{green} dot, and incorrect strategies in \textcolor{red}{red}.

We deemed the simplification of the model necessary to allow a timely execution of the test series, which contained 400 experiments in total (100 for each grid resolution). 
Detailed information about the time of the evaluation is shared at the end of the section.

As expected, a smaller cumulative contamination probability leads the strategy to de-prioritise cleaning the rooms regularly, since the probability of them being contaminated stays low.
This, at some point, causes the strategy to not satisfy the recurrence criteria.
Similarly, a lower value of $\mathit{a\_bit}$ causes the strategy to
prioritise saving battery over resetting the contamination flags,
which reduces the energy consumption significantly, but at some point
at the cost of strategy correctness.

In the evaluation, the optimal strategy per grid resolution uses less or equal energy with a larger grid resolution (84, 64, 64, and 36 for $g$ of 1, 2, 3, and 4, respectively).
There is also a difference in the number of distinct strategies that are generated between the different $g$, where the experiment with a resolution of 4 generates more unique strategies than the experiments with lower grid resolutions, where many parameters lead to the same generated strategy.

Finally, we can observe distinct areas of incorrect, non-recurrent and
recurrent strategies that depend on these parameters, where the
optimal recurrent strategy lies on the border between recurrent and
non-recurrent strategies.  While the states in $\omega$ are
ordered for the choice of a worst case, the border area is at best an
approximation of a Pareto front, the non-convex reward function defined
by $R$ combined with the belief-MDP approximation $B(M)$ may
lead to optimal strategies remaining hidden from the search.

Beyond the reward structure $R^{utilisation}$ used for strategy
pre-selection, checking the PLTL safety property UT
(\Cref{tab:strat-req}) ensures that the finally chosen strategy only
cleans outside the room utilisation schedule (\Cref{room-times}).

\paragraph{Some Key Data.}

The experiments were conducted on an AMD FX(tm)-8350 Eight-Core
Processor with 32\,GiB of RAM  running Ubuntu 22.04.4 LTS. However,
\textsc{Prism}\xspace was restricted to one core and 12\,GiB of RAM. 
The reduced model in~\Cref{sec:experiment-parameter} contains 4879
states and 35014 transitions, while the reduced model $M^\sigma$
contains 23 states and 23 transitions.  The verification consists of 13 PLTL formulas containing 25 propositions.
We ran the experiments with parameters of $m=4$, $\mathcal{R}_i.pr = 0.02, 0.04, \ldots, 0.16$ and a\_bit = $1, 3, 6, 10, 17, 32, 100, 316, 1000, 3162$. The \emph{cumulative probability} can be derived from $\mathcal{R}_i$: $\sum_{i=1,\ldots,m} \mathcal{R}_i.pr = 0.08, 0.16, \ldots, 0.64$.
The strategy synthesis took about 5, 16, 60, and 200 seconds for a grid resolution of 1, 2, 3, and 4, respectively, while the verification took about one second for a given~$\omega$.
Evaluating the entire test series took about 8 hours sequentially,
although this process could be easily parallelised.

\section{Discussion}
\label{sec:discussion}

\paragraph{Selecting the Recurrence Area $\omega$.}

When evaluating the strategy, $\omega$ was chosen either by
examining the strategy manually (\Cref{fig:run}) or by verifying a
list of probable $\omega$s.  Further work may focus on finding
probable $\omega$s from the room layout, and generating strategies
which fulfil these $\omega$s.

\paragraph{Complexity of the Cleaning Scenario.}

For the evaluation, we considered a rather simple room layout.
The performance of the above described method may be different with
larger room graphs, more complex room layouts, a larger number of
robots, and tighter restrictions on battery charge and room
utilisation.

Moreover, our model allows us to find strategies that operate with a
varying number of robots.  Given that some robots remain idling all
the time, our optimal synthesis could also be used to find the
smallest subset $B_{\min}\subseteq B$ or minimal number
$k_{\min}\leq k$ of robots for an optimal task performance.

The complexity of the model is heavily dependant on the number of
rooms, the maximum time $T$, and the maximum charge of the robots.
Following the comprehensive scheme in \Cref{sec:experiment-parameter},
we were able to calculate a 12-hour ($T=12$) cleaning schedule for 3
robots with 11 rooms and a maximum charge of 6 in 15 hours.  The
corresponding belief-MDP $B(M)$ contains $\approx$\,690k states
and $\approx$\,11.8m transitions.

\paragraph{Adjusting Grid-Resolution vs.\ Filtering Strategies.}

For industry-size POMDPs, a high resolution $g$ can
lead to an impractically high computational effort when solving the
mostly NP-hard approximate analysis (i.e.,\xspace verification, synthesis)
problems.  Hence, our approach is to keep $g$ just fine enough to find
some (not necessarily globally optimal) strategy $\sigma$ and verify
more nuanced properties of the quasi-MDP\footnote{non-probabilistic,
  deterministic, with full observability} $M^\sigma$ derived from
$M$ by applying $\sigma$.  In $M^\sigma$, verification is
simpler (no belief-MDP $B(M)$ is computed), also the
strategy (integrated in $M^\sigma$) can directly observe the outcome
of each action and does not have to memorise a finite observation
history.
Despite the expansion of $\mathcal{R}_i.d$ to integers, $M^\sigma$'s state space is
expected to be smaller than $B(M)$'s state space for the applied values of $g$.

\paragraph{Parameter Selection.}

For the evaluation in \Cref{sec:experiment-parameter}, a set of values
for the parameters \emph{a\_bit} and the \emph{contamination
  probability} was chosen.  Via~$g$ (\Cref{sec:orgae569ee}), we
reduced the resolution of the fixed grid~(i.e.,\xspace a wider grid width) to
limit the number of states in the belief space
approximation~$B(M)$.
However, our findings in \Cref{sec:experiment-parameter} suggest that
increasing the resolution, while keeping the cumulative contamination
probability around 40\,\% and the value of \emph{a\_bit} around 300 leads to the synthesis of better
strategies.  However, these values may not be universally favourable
for any room layout, and it may be possible to synthesise better
strategies using a different set of parameters.  Further work may
focus on better ways of parameter selection.

\paragraph{Generalisation to Other Applications.}

The running example in our case study focuses on a cleaning robot
collective.  However, we think that our approach and model can be
transferred rather straightforwardly to other spatio-temporal settings
with recurrent tasks, for example,
\begin{itemize}
\item firefighting drone collectives tasked with repetitive
  sector-wise fire detection and water distribution and with partially
  observable quantities such as ground temperature and extinction
  level;
\item geriatric care robot collectives tasked with recurrent
  monitoring and care-taking tasks (e.g.,\xspace medication supply) with
  patient satisfaction and health status being partially observable;
\item general patrolling collectives tasked with monitoring or
  supervising specific
  environments~\cite{Portugal2011-SurveyMultirobot}.
\end{itemize}

\section{Conclusion}
\label{sec:conclusion}

We proposed an approach using weighted, partially observable
stochastic models (i.e.,\xspace reward{\Hyphdash}enhanced POMDPs) and
strategy synthesis for optimally coordinating tasked robot collectives
while providing recurrence and safety guarantees on the resulting
strategies under uncertainty.  Along with that, we discussed guidance
on POMDP modelling and strategy synthesis.  We focused on a
cleaning robot scenario for public buildings, such as schools.
Our notion of \emph{correctness} combines (i) safe recurrence (e.g.,\xspace
repetitively accomplish the cleaning task while avoiding to collect
penalties), (ii) robustness (e.g.,\xspace correctness under worst-case
contamination), and (iii) optimality (e.g.,\xspace minimal energy
consumption).

For scaling up strategy synthesis to scenarios beyond what can easily
be tackled by stochastic game-based synthesis, we addressed the key
challenge~\cite{Gleirscher2023-ManifestoApplicableFormal} of reducing
the state space and the transition relation of a na\"ive model via
partial observability (hiding details of stochastic room
contamination) and by employing simultaneous composition (for robot
movement).  \textsc{Prism}\xspace's grid-based POMDP approximation allowed us
to adjust the level of detail of the belief space to synthesise
strategies more efficiently.
Furthermore, we softly encode the strategy search space
using penalties and optimisation rewards and can, thus, shift the
verification of more complicated properties to a later stage working
with an unweighted and non-probabilistic behavioural model, again
using a more detailed, numerical state and action space.
However, decoupling synthesis from verification can require
time-consuming experiments~(\Cref{sec:experiment-parameter}) to
identify regions of the parameter space for ensuring the existence of
good recurrent strategies.

In future work, we will improve finding $\omega$ ensuring the
existence of correct strategies~(i.e.,\xspace green dots in \Cref{runs}).
Ideally, we avoid defining $\omega$ explicitly~(e.g.,\xspace by hiding
time).  In a larger example, we want to allow invariant-narrowing with
$\phi_{\mathsf{r}}$ and observable stochasticity in the environment, such
that $\sigma$ can depend on arbitrary variables.  The reset of the
contamination flag on a room visit~(Db) could be refined by a
decontamination rate in $M^\sigma$.
Moreover, we aim to use multi-objective queries to include further
criteria~(e.g.,\xspace minimal contamination) for Pareto-optimal strategy
choice.  While \textsc{Prism}\xspace imposes some limits on the combination of
queries and constraints, we will need to see how we can use tools such
as \textsc{EvoChecker}~(as, e.g.,\xspace used in
\cite{Vazquez2022-SchedulingMissionsConstrained}) for POMDPs.
Also, we can further reduce the action set by taking into account
trajectory intersections in the simultaneous movements
(cf.~\Cref{fig:bot-coordinator}).
Finally, we want to connect the synthesis pipeline with code
generation, such as demonstrated in our previous
work~\cite{Gleirscher2022-VerifiedSynthesisSafety}.

\bibliographystyle{eptcs}
\bibliography{main}
\end{document}